\documentclass[pra,twocolumn,titlepage,nofootinbib,amsmath,showpacs]{revtex4}%
\usepackage{amsmath}
\usepackage{amsfonts}
\usepackage{amssymb}
\usepackage{graphicx}%
\begin{document}
\title{Precision physics of simple atoms and constraints on a light boson with ultraweak coupling}
\author{S.~G.~Karshenboim}
\email{savely.karshenboim@mpq.mpg.de} \affiliation{D.~I. Mendeleev
Institute for Metrology, St.Petersburg, 190005, Russia
\\ {\rm and}
Max-Planck-Institut f\"ur Quantenoptik, Garching,
85748, Germany}

\begin{abstract}
Constraint on spin-dependent and spin-independent Yukawa potential
at atomic scale is developed. That covers constraints on a coupling
constant of an additional photon $\gamma^*$ and a pseudovector
boson. The mass range considered is from $1\;{\rm  eV}/c^2$ to
$1\;{\rm MeV}/c^2$. The strongest constraint on a coupling constant
$\alpha^\prime$ is at the level of a few parts in $10^{13}$ (for
$\gamma^*$) and below one part in $10^{16}$ (for a pseudovector)
corresponding to mass below $1\;{\rm keV}/c^2$. The constraints are
derived from low-energy tests of quantum electrodynamics and are
based on spectroscopic data on light hydrogen-like atoms and
experiments with magnetic moments of leptons and light nuclei.
\end{abstract}
\maketitle

\section{Introduction}

Precision physics of simple atoms is a field, which provides us
with values of various fundamental constants with high accuracy
\cite{codata2006} and enables to perform low-energy tests of quantum
electrodynamics (QED) \cite{my_rep}. Indeed, it is important to
verify specific theoretical calculations, which are quite advanced
and sophisticated \cite{EGS}, although we hardly have any doubts about
QED as a universal framework to describe kinematics of photons,
theory of electromagnetic interactions of point-like particles and
phenomenology of electromagnetic interactions of structured objects,
such as hadrons.

Meantime, various unification theories suggest new particles, which
have not yet been observed (see, e.g., \cite{holdom,pospelov}). One
class of such particles deals with light [electrically] neutral
particles with ultraweak coupling to conventional matter. Stable
neutral particles of this kind are also a candidate for the dark
matter \cite{cosmo}. That means that, while energetically such
particles are reachable in existing particle-physics experiments,
their ultraweak coupling makes their production and detection rate
so low that their observation is in reality impossible.

The atomic physics allows one, on the contrary, very accurate
measurements, and thus certain constraints within a keV/$c^2$ range
of mass of an intermediate boson are in reach. Below we consider two
basic options. At first, we constrain a kind of an additional photon
$\gamma^*$, which interacts universally with all charged particles.
The other kind of particles we study is a pseudovector boson, which
induces a spin-dependent interaction between atomic constituents.

Dealing with effects at atomic scale, it is more advantageous to
apply the coordinate-space consideration and to constrain a certain
long-range interaction in the form of the Yukawa-potential
correction to the Coulomb law
 \begin{eqnarray}
 -\frac{\alpha}{r} &\to&-\frac{\alpha_{\rm eff}(r)}{r}=-
 \frac{\alpha+\alpha^\prime e^{-\lambda r}}{r}\;,
 \label{ar1}\\
 -\frac{\alpha}{r} &\to& -\frac{\alpha+\alpha^{\prime\prime}
 \bigl({\bf s}_1\cdot{\bf s}_{2}\bigr)e^{-\lambda
 r}}{r}\label{ar2}\;,
 \end{eqnarray}
which is multiplied by the nuclear charge $Z$ if necessary. Here,
the first line describes $\gamma^*$, while the second line is for an
axial boson and ${\bf s}$ stands for the spin of a particle.

A requirement to constrain the Yukawa terms in  a certain range of
$\lambda$ is quite simple: it is necessary to compare two
experiments, which have different sensitivity to the correcting
term. Here we compare experiments, which involve different
distances. While in Sects.~\ref{s:gam} and \ref{s:1} we deal with
distance scales, different by orders of magnitude, in the
consideration in Sect.~\ref{s:12} the compared distances are
relatively close, but the sensitivity is enhanced.


\section{Constraining an extra photon $\gamma^*$\label{s:gam}}

The most intensively studied atomic scale is at a few values of the
Bohr radius, $a_0\simeq 0.53\times10^{-10}\;$m. For the Yukawa
radius, equal to $a_0$, the related mass $\lambda$ of the
intermediate particle is $3.5\;$keV. (We apply here relativistic
units, in which $\hbar=c=1$.)

The fundamental constants related to this scale are
\begin{eqnarray}
R_\infty&=&10\,973\,731.568\,527(73)\;{\rm m}^{-1}\;,\label{rya}\\
\alpha^{-1}&=&137.035\,999\,59(53)\label{aa}\;.
\end{eqnarray}

Before applying both values to constrain $\alpha^\prime$, we briefly
explain their origin. Once we suggest a substitution (\ref{ar1}),
both results become related to $\alpha_{\rm eff}(r)$ at a
certain effective value $r_{\rm eff}\sim (1-4)a_0$.

The value (\ref{rya}) is from the evaluation \cite{codata2006} of
experimental and theoretical results, where a statistically
dominant contribution involves data on the $1s$ and $2s$ states
\cite{mpq,mpq_d,paris}, while the Yukawa correction (\ref{ar1}) to
other excited levels, also involved, is of marginal importance.

The value (\ref{aa}) is obtained by combining (\ref{rya}) with
values of $h/M$ for caesium \cite{chu} and rubidium \cite{rb} and
with various results on auxiliary data (see \cite{codata2006} for
details).

Meantime, certain results related to other mass/distance scales are
also available. For longer distances one can apply the value
\cite{MIT}
\begin{equation}
R_\infty=10\,973\,731.568\,34 (69)\;{\rm m}^{-1}
\end{equation}
obtained from a transition between the circular states (i.e., states
where $l=n-1$) with $n=27-30$ in the hydrogen atom and thus related
to distances of about $10^3a_0$. The uncertainty above is tripled
against its original value \cite{MIT} because the result is a
preliminary one. Nevertheless, as it was confirmed by authors of the
experiment \cite{kleppner}, that is rather an overconservative
estimation. We have to remark that while this uncompleted experiment
is rather of marginal importance in determination of the Rybderg
constant (see, e.g., \cite{codata2006}), it is crucial to constrain
various exotic effects at distances of $10^{-7}-10^{-6}\;$m. There
is no other compatible experimental data at this scale.

The fine structure constant
\begin{equation}
\alpha_{g\!-\!2}^{-1}=137.035\,999\,084(51)\;,
\end{equation}
derived from  the anomalous magnetic moment of an electron by
combining the experimental result \cite{aexp} with theory
\cite{ath}, represents physics of shorter distances, comparable with
the Compton wavelength of the electron  $\sim \lambdabar_C
=a_0/\alpha=1/m_e$. (Indeed, instead of a substitution (\ref{ar1}),
for the evaluation of the related correction one has to apply a
complete propagator of $\gamma^*$.)

\begin{table}[htbp]
 \begin{center}
 \begin{tabular}{cccc}
 \hline
~~~~~&Mass range  & $r$ &$\alpha(r)-\alpha (a_0)$ \\[0.8ex]
 \hline
{\em a} & $4\;{\rm eV}\ll\lambda\ll1\;{\rm keV}$& $10^3a_0$
&$\bigl(0.6\pm2.3\bigr)\times 10^{-13}$ \\[0.8ex]
{\em b} &$4\;{\rm keV}\ll\lambda\ll0.5\;{\rm MeV}$& $\lambdabar_C$ &$\bigl(2.7\pm2.9\bigr)\times 10^{-11}$\\[0.8ex]
\hline
 \end{tabular}
\caption{The constraint on the deviation of the effective long-range
interaction $\alpha(r)/r$ from the Coulomb exchange due to possible
presence of $\gamma^*$. Here,
$\alpha(r)=\alpha(\infty)+\alpha^\prime \exp(-\lambda r)$ and $r$ is
a characteristic distance to be compared with $a_0$. The related
distance range is $\lambda^{-1}=0.5\times10^{-7}\;$m (for
$\lambda=4\;$eV), $0.5\times10^{-10}\;$m (for $\lambda=4\;$keV) and
$0.4\times10^{-12}\;$m (for $\lambda=0.5\;$MeV).
 \label{t:sum}}
 \end{center}
 \end{table}

\begin{figure}[thbp]
\begin{center}
\resizebox{0.8\columnwidth}{!}{\includegraphics{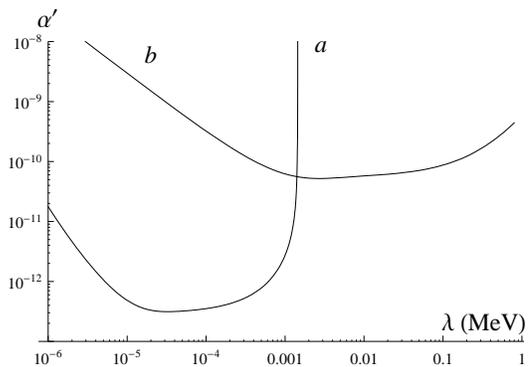}}
\end{center}
\caption{Constraints on a long-range spin-independent interaction
from hydrogen spectroscopy and $g_e\!-\!2$, including a constraint
from hydrogen spectroscopy with low and high ($n\simeq30$) Rydberg
states ($a$), a comparison of the low states and $g_e\!-\!2$ ($b$).
The lines are for the upper bound on $|\alpha^\prime|$ and the
confidence level corresponds to one standard deviation.}
\label{f:ry}       
\end{figure}

Comparison of $\alpha_{\rm }(a_0)$ with results at other scales
delivers us constraints on $\alpha^\prime$. The results for
asymptotic areas (where the strong inequalities on $\lambda$ hold)
are summarized in Table~\ref{t:sum}, while the constraints in
Fig.~\ref{f:ry} are also applicable for an intermediate area
$\lambda\sim 1\;$keV as well as on the edges of the considered
region ($\sim 1\;$eV and $\sim 1\;$MeV). We also note that to
saturate limits for the region, defined by a strong inequality, it is
sufficient to have $\lambda$ larger/smaller than the related value by
a factor of $3-5$.

To interpret constraints in Fig.~\ref{t:sum}, we note that the $a$
constraint is related to a correction (\ref{ar1}) for the Coulomb
interaction of an electron and proton, while the $b$ constraint is
from a comparison of an electron-proton long-range interaction at
atomic scale and electron-electron interaction at scale of
$\lambdabar_C$.

In principle, one could have in mind not $\gamma^*$, but a light
intermediate particle, which interacts in a different way with
various charged particles. It is possible to proceed further for
this case and this option will be explored in details elsewhere.

The constraints can be also applied for a non-vector intermediate
particle coupled both to protons and electrons. A potential, which
has a static spin-independent interaction (i.e., the Coulomb-like
component), satisfies (\ref{ar1}) and is directly applicable in
atomic calculations for $R_\infty$ at distances of both important
scales, $a_0$ and $10^3a_0$. That is related to true scalar, vector,
tensor etc. intermediate particles and the constraint $a$ in
Table~\ref{t:sum} and Fig.\ref{f:ry} stands for them. On the
contrary, pseudoscalars, vectors etc. produce various spin-dependent
potentials, that does not affect any determination of $R_\infty$.

Meantime, a calculation of corrections to $g_e\!-\!2$ with all kinds
of intermediate particles, coupled to electrons, produces results of
the same order of magnitude. Suggesting that a coupling of the boson
to electrons and to protons is the same, we can consider the $b$
constraint as a rough estimation for true scalars. For pseudoscalars
and vectors, this rough estimation is now for a pure boson-electron
coupling and has an interval of applicability, extended to longer
distances.

\section{Constraining a pseudovector boson from
the $1s$ HFS interval\label{s:1}}

Until  very recently, it was the $1s$ hyperfine structure (HFS)
interval in the hydrogen atom that was the most accurately measured
quantity in general and the most accurately measured quantity
related to a simple atom in particular. However, its application to
fundamental problems used to be limited because of uncertainties
related to the proton structure. In this section in particular we
derive from the hydrogen HFS a constraint on a pseudoscalar
particle, which is much stronger than a constraint on $\gamma^*$.
Still a constraint from muonium HFS, which is free of problems with
the nuclear structure, is even stronger.

The HFS in two-body atoms opens a possibility to test a
spin-dependent long-range interaction and to look for the Yukawa
term in (\ref{ar2}). The ground state HFS interval is known with
high accuracy for a few light hydrogen-like atoms (see \cite{my_rep}
for details and references).

A constraint here is based on the fact that the leading contribution
to the HFS interval can be obtained from the known value of the nuclear
magnetic moment. The latter is determined for a number of nuclei
from experiments on behavior of a bound nuclear magnetic moment at a
macroscopic magnetic field (see \cite{codata2006} for details and
references). In principle, 
only some of these measurements are sensitive to the Yukawa term in
(\ref{ar2}), and the involved macroscopic distances vary in a broad
range. Summarizing, we can definitely conclude that for the Yukawa
radius below, say, $l_0=1\;$cm, the results for magnetic moments are
not affected by the correcting term.

Meantime, for $a_0\ll \lambda^{-1}\ll l_0$ the HFS interval in
muonium and hydrogen is shifted by $
({-\alpha^{\prime\prime}}/{\alpha})Z^2R_\infty$.
The  strongest constraint
\begin{equation}\label{muonium}
\alpha^{\prime\prime}= \bigl(1.6\pm 6.0\bigr)\times 10^{-16}
\end{equation}
is derived from muonium physics \cite{mu1shfs}, while the results,
involving other light two-body atoms, such as
$\alpha^{\prime\prime}({\rm H})= \pm 1.6\times 10^{-15}$ and
$\alpha^{\prime\prime}({\rm D})= \pm 8\times 10^{-15}$,
are somewhat weaker and, in fact, less reliable because of
uncertainties in theoretical understanding of the nuclear
contributions (see, e.g., \cite{d21th,my_rep}).

\begin{figure}[thbp]
\begin{center}
\resizebox{0.95\columnwidth}{!}{\includegraphics{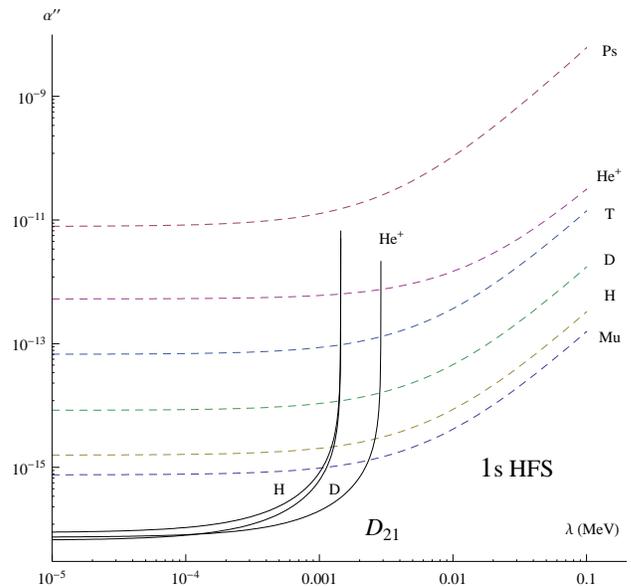}}
\end{center}
\caption{Constraint on a long-range spin-dependent interaction from
HFS intervals of two-body atoms. The dashed lines are for the upper
bound on $|\alpha^{\prime\prime}|$ derived from the $1s$ HFS data
for muonium, positronium, hydrogen, deuterium, tritium and helium-3
ion. The solid lines present constraints from data on $D_{21}$ in
hydrogen, deuterium and helium-3 ion. The confidence level
corresponds to one standard deviation.}
\label{f:hfs}       
\end{figure}

The constraints, obtained from all available precision data on the
$1s$ HFS in light atoms are summarized in Fig.~\ref{f:hfs}. They are
extended there to shorter distances ($\lambda^{-1}<a_0$).

The constraints also include the one from positronium experiments,
obtained in a different approach, through comparison of
spin-dependent (HFS interval) and spin-independent (determination of
$e$ and $m_e$) physics.

\section{Constraining a pseudovector
 by comparing
the $1s$ and $2s$ HFS intervals
\label{s:12}}

Summarizing, the constraints from the $1s$ HFS are limited either by
the uncertainty of calculations of nuclear effects (for atoms with
hadronic nuclei) or by experimental accuracy in leptonic atoms.
Since both uncertainties in fractional units are larger than
experimental uncertainty of the $1s$ HFS interval in hydrogen and
some other atoms, the high accuracy of measuring the HFS intervals
is not completely utilized and progress is possible.

As was shown in \cite{d21th}, the specific difference
\begin{equation}\label{defd21}
D_{21}=8\times E_{\rm hfs}(2s)-E_{\rm hfs}(1s)\;,
\end{equation}
is essentially free of such a problem. The uncertainty of the
nuclear effects is very much reduced and the accuracy of the $2s$
interval, being lower than for the $1s$, is still higher than in
experiments with unstable leptonic atoms, such as muonium and
positronium.

To get use of the difference, accurate experimental data on the $2s$
HFS interval are required.
Data with appropriate accuracy are available for few atoms: for
hydrogen \cite{exph2s}, deuterium \cite{expd2s} and helium-3 ion
\cite{exphe2s}. The related theory is reviewed in \cite{d21th1}.

The Yukawa correction in (\ref{ar2}) generates a correction for the
HFS interval of the $ns$ state, which is proportional to $n^{-2}$
for $\lambda^{-1}\gg a_0$ and thus does not vanish in
(\ref{defd21}). That allows to set a constraint on the
spin-dependent Yukawa term (see Table~\ref{T:exp} and
Fig.~\ref{f:hfs}).

\begin{table}[phtb]
\begin{tabular}{clll}
\hline
Atom & ~Experiment~ & ~~~~~Theory~~~ & ~~~~~~~~~~$\alpha^{\prime\prime}$  \\
 & ~~~~~[kHz] & ~~~~~~[kHz]  &    \\
 \hline
H & ~~~~~~48.923(54) & ~~~~~48.953(3)  & $~~\bigl(3.3\pm5.9\bigr)\times10^{-17}$ \\
D & ~~~~~~11.280(56) & ~~~~~11.3125(5) &  $~~\bigl(2.4\pm4.1\bigr)\times10^{-17}$ \\
$^3$He$^+$ &$-1189.979(71)$ &$-1190.08(15)$ & $\bigl(-2.8\pm4.6\bigr)\times10^{-17}$\\
\hline
\end{tabular}
\caption{Comparison of experiment and theory for the $D_{21}$ value in
light hydrogen-like atoms. The constraint on $\alpha^\prime$ is
related to $\lambda\ll 1\;$keV.\label{T:exp}}
\end{table}

Comparing the HFS constraints in Fig~\ref{f:hfs}, we note that the
$D_{21}$ constraints are stronger for $\lambda\leq 1\;$keV, while
the $1s$ HFS constraint for $\alpha^{\prime\prime}$ is stronger in
the case of a heavier mass range.

\section{Summary}

Concluding, we have demonstrated that precision physics of simple
atoms is a powerful tool to constrain vector and pseudovector
particles, coupled to leptons and nuclei, with a preferred mass
range of keV/$c^2$. The latter in many cases can be extended into
one-MeV domain. The stability of the intermediate particle is not
required and it is sufficient for our consideration that the width
is substantially smaller than the mass ($\tau^{-1}\ll \lambda$).


The width of the boson, $\tau^{-1}$, can be in particular induced by
the interaction with the charged particles, which is constrained in
this paper. However, that is not the only possible mechanism. On the
contrary, it may happen that there is an interaction of the
intermediate boson with the photons, which induces the interaction
with charged particles. (The situation with $\pi^0$ and $a_1$
exchange, studied while calculating the hadronic contributions to
the muonium HFS \cite{lbl}, is similar---the interaction of both
mesons with photons couples them to electrons and muons.) That may
provide smaller values of the lifetime of the intermediate particle.

Here, we have considered an option of a light boson with an ultraweak
coupling, which has not been well explored in particle physics (see,
e.g., \cite{pdg}).

Various constraints on long-distance interactions of this range
could also come from experiments, studying Casimir effect at small
distances \cite{casimir}. However, here, we consider somewhat
smaller distances, which are not accessible in those experiments.
Yukawa potentials, related to $\gamma^*$ or to a pseudovector meson,
are also outside of reach in those investigations, which involve a
long-range spin-independent interaction of bulk neutral matter. So,
our constraints are complementary to Casimir-effect studies.

The keV mass range can be explored by means of astrophysics and
cosmology \cite{astro,cosmo}. Those constraints involve additional
details such as the lifetime and other couplings and they are also
complementary to our constraint for $\alpha^\prime(\lambda)$.

The novel method suggested here, based on a study of a few specific
atomic transitions understood theoretically and experimentally with
extremely high accuracy, covers an area of parameters
(mass--coupling constant), not available by other methods (at least
in a model-independent way). The details of our evaluations are to
be published elsewhere. We expect that some other atomic data can be
also useful for additional constraints.

This work was supported in part by RFBR (grants \#\# 08-02-91969 \&
08-02-13516) and DFG (grant GZ 436 RUS 113/769/0-3). The author is
grateful to Dan Kleppner, Andrej Afanasev, Vladimir Korobov, Masaki
Hori, Astrid Lambrecht, Dmitry Toporkov, Eugene Korzinin, Simon
Eidelman, and Maxim Pospelov for useful and stimulating discussions.

\end{document}